\documentclass[10pts,showpacs,preprint,aps]{revtex4}
\usepackage{graphicx}
\usepackage{dcolumn}
\usepackage{bm}
\usepackage{amssymb}
\usepackage{amsmath}
\usepackage{mathrsfs}
\begin{document}
\setcounter{page}{1}
\vskip 2cm
\title
{Path-integral derivation of black-hole radiance inside the de-Rham-Gabadadze-Tolley formulation of massive gravity}
\author
{Ivan Arraut}
\affiliation{State Key Laboratory of Theoretical Physics, Institute of Theoretical Physics,
Chinese Academy of Science, Beijing 100190, China}

\begin{abstract}
If we apply the path integral formulation in order to analyze the particle creation process of black-holes inside the non-linear formulation of massive gravity, it is possible to demonstrate that the effect of the extra-degrees of freedom is to deform the periodicity of the poles of the propagator in the complex $t$-plane. This might create the effect of extra-particle creation process at scales where the extra-degrees of freedom become relevant. For stationary solutions, depending on the values taken by the free parameters of the theory, the periodicity structure of the propagator reveal two effects. The first one is a shift on the positions of the pole of the propagator with respect to the GR case, affecting then the instant at which the particles are detected. The second one is the existence of branch points, affecting then the perception of particles. The branch point can be finite (including the zero order case) or infinite depending on the free-parameters of the theory.
\end{abstract}
\maketitle 

\section{Introduction}
In \cite{1, On the app}, the first analysis of particle creation process of black-holes in dRGT massive gravity was performed. The analysis done in \cite{1} used the Hawking method studied in \cite{Hawking}, but in an implicit way. This means that the extra-degrees of freedom were not analyzed explicitly, but rather they appeared implicitly inside the definitions of extended coordinates (St\"uckelberg functions). This method just take the advantage of the fact that the extra-degrees of freedom appear in a similar way as the coordinate transformations in general relativity (GR). This is just the well known St\"uckelberg trick applied at the non-linear level. In \cite{On the app}, the path integral formulation was explored. In such a case, the analysis of the periodicity of the poles of the propagator revealed the possibility of an extra component of radiation coming from the fact that the extra-degrees of freedom of the theory reproduce a vacuum degeneracy. Different vacuums will naturally define different values of the Hawking radiation. This extra-component of radiation, is relevant at scales where the extra-degrees of freedom of the theory are important because they create a distortion on the definition of time. If we take the origin of coordinates on the source, this happens for distances larger than the Vainshtein radius $r_V$ and only observers defining the time in an arbitrary direction with respect to the St\"uckelberg function ($T_0(r,t)$) are able to detect it since they naturally define different notions of vacuum. The observers defining the time in the direction of $T_0(r,t)$ will describe a vacuum as in the GR case. For such privileged observers, the Hawking radiation happens to be as in the standard case. Although the result found in \cite{On the app} is general, in such a case the author did not evaluate explicitly the new value of temperature. This value will depend on how the periodicity of the propagator is affected by the presence of the extra-degrees of freedom. 
The calculations performed by using the path integral method, are simplified if we work by defining the St\"uckelberg functions in agreement with the extended versions of the advanced and retarded coordinates as has been done in \cite{1, On the app}. In the extended version of the coordinates, the St\"uckelberg function denoted by $T_0(r,t)$ appears in replacement of the time-coordinate. In this manuscript, I study the particle creation process of black-holes by using the path-integral formulation. I explore the analyticity properties of the propagator inside the non-linear formulation of massive gravity proposed by de-Rham, Gabadadze and Tolley (dRGT). The general result suggests that for St\"uckelberg functions given by $T_0(r,t)=St+A(r,t)$, with $A(r,t)$ having some non-linear time-dependence, the periodicity properties of the poles of the propagator change and the effect of the extra-particle creation for the observers defining different notions of time is evident. The dynamical metric in this case is not necessarily stationary (although the non-stationary condition can appear at the perturbative level). However, if $A(r,t)$ has a linear time-dependence, then the properties of the poles of the propagator are also affected and the stationary condition of the dynamical metric would be satisfied. These general situations are not explored in this manuscript. Here I concentrate the analysis on the black-hole solution found in \cite{Kodama} for the case $\beta=(3/4)\alpha^2$, with $\beta$ and $\alpha$ representing two free parameters of the theory. This combination of parameters is special because it provides a solution with zero cosmological constant ($\Lambda$) and a non-zero graviton mass. Then it is easier to analyze the effect of the extra-degrees of freedom at large scales. Similar solution was found in \cite{gabagabaga}. The stationary condition of this solution, guarantees that the St\"uckelberg function is linear in time and given by $T_0(r,t)=St+A(r)$, with $A(r)$ representing a spatial dependent function carrying the information of the extra-degrees of freedom. Here $A(r)$ is time-independent. Then after analytical extension of the time-coordinate $t$, the effect of the function $A(r)$, is to create a shift effect for the poles of the propagators if we compare them with respect to the GR case. In addition, if we express the extended version of the Kruskal coordinates as an explicit function of time $t$, then after the analytical extension of the time-coordinate, the periodicity of the poles of the propagator is given by $8\pi M/S^2$, with $S$ depending on the free-parameters of the theory. The relation $M/S^2$ marks the possibility of having a branch point condition. The order of the branch point is given by $1/S^2$ and it depends on the value taken by the parameters of the theory. For the case $\beta=(3/4)\alpha^2$, the order of the branch point becomes zero when $\alpha\to-2$. It is infinite for $\alpha=0$ and in general is finite for other values taken by the parameter $\alpha$. If $S=1$ (when $\alpha\to\pm\infty$), the periodicity of the propagator is the same as in the GR case, however the shift of the poles of the propagators produced by the function $A(r)$ in this particular case, although does not affect the value of temperature with respect to the GR case (for the case $S=1$), it affects the space-time locations where the observers detect particles. In other words, if an observer moving under the theory of GR travels through a line of constant $r, \theta, \phi$, then he will disagree with the instants (of time) at which the particles appear if he compare his results with the observers describing the physics in agreement with the dRGT theory of massive gravity. This is equivalent to a time-dilation effect reproduced by the extra-degrees of freedom at large scales. In some sense, this is expected because the effect of the extra-degrees of freedom is to reproduce a vacuum degeneracy and the definition of vacuum depends on the time coordinate. For the case of the branch points, when they become of order zero or infinite order, the effect of particle creation coming from the black-hole is completely suppressed for the observers describing this kind of solutions. The paper is organized as follows: In Section (\ref{eq:Final1}), I make a brief review of the Schwarzschild de-Sitter solution inside the non-linear formulation of massive gravity; this solution is discussed in detail in \cite{Kodama}. In Section (\ref{eq:pathspart}), I make a brief review of the path integral formulation applied to the analysis of the particle creation process of black-holes. This analysis was originally proposed by Hartle and Hawking \cite{Hartle}. Section (\ref{eq:non-linearStu}) is dedicated to explore the non-linear version of the St\"uckelberg trick. The idea is to explain the distinction between coordinate transformations and the St\"uckelberg trick. From the perspective of GR, the St\"uckelberg trick looks like a dipheomorphism transformation. Inside dRGT massive gravity, this is not the case. In Section (\ref{eq:Analitycity}), I explore the analyticity properties of the propagator. This is the key section of the manuscript. In Section (\ref{eq:Blackrad}), I explain the expected modifications for the black-hole radiance due to the extra-degrees of freedom of the theory for the case $\beta=(3/4)\alpha^2$. This section explores the connection between the rates of emission and absorption. In Sec. (\ref{Secla}), I make a brief comparison between the standard case of massive gravity analyzed in this paper and the more explicit situation where the fiducial metric really couples to matter. Finally, in Section (\ref{eq:Con}), I conclude.

\section{The Schwarzschild de-Sitter solution in dRGT: Unitary gauge} \label{eq:Final1}

In \cite{Kodama}, the S-dS solution can be written explicitly as:

\begin{equation}
ds^2=G_{tt}dt^2+G_{rr}S^2dr^2+G_{rt}(drdt+dtdr)+S^2r^2d\Omega_2^2,
\end{equation}
where:

\begin{eqnarray}   \label{eq:drgt metric}
G_{tt}=-f(Sr)(\partial_tT_0(r,t))^2,\;\;\;\;\;G_{rr}=-f(Sr)(\partial_rT_0(r,t))^2+\frac{1}{f(Sr)},\;\;\;\;\;\nonumber\\
G_{tr}=-f(Sr)\partial_tT_0(r,t)\partial_rT_0(r,t),
\end{eqnarray}
and $f(Sr)=1-\frac{2GM}{Sr}-\frac{1}{3}\Lambda (Sr)^2$, with $S=\frac{\alpha}{\alpha+1}$ being the scale factor which depends on the free-parameters of the theory. In this previous solution, all the degrees of freedom are inside the dynamical metric. The fiducial metric in this case is just the Minkowskian one given explicitly as:

\begin{equation}   \label{eq:drgt metric223}
f_{\mu\nu}dx^\mu dx^\nu=-dt^2+dr^2+r^2(d\theta^2+r^2sin^2\theta).
\end{equation}
 The solution (\ref{eq:drgt metric}) can be equivalently written as:

\begin{equation}   \label{eq:drgt metric2}
ds^2=-f(Sr)dT_0(r,t)^2+\frac{S^2dr^2}{f(Sr)}+S^2r^2d\Omega^2,
\end{equation}
where $T_0(r,t)$ corresponds to the St\"uckelberg function. This function contains the information of the extra degrees of freedom  in agreement with the formulation introduced in \cite{1,K}. The metric (\ref{eq:drgt metric2}) is then diffeomorphism invariant \cite{1, K}. The gravitational degrees of freedom of the St\"uckelberg fields appear inside $T_0(r,t)$ through the spatial (temporal) dependence of the function. For the family of solutions with two free-parameters and the St\"uckelberg function constrained as has been found in \cite{Kodama}, the cosmological constant is given by \cite{Kodama}:

\begin{equation}
\Lambda=-m^2\left(1-\frac{1}{S}\right)\left(2+\alpha-\frac{\alpha}{S}\right).
\end{equation}
This constant is zero when the two free-parameters of the theory $\alpha$ and $\beta$ satisfy the relation $\beta=(3/4)\alpha^2$. Independent of the relation between the two free parameters of the theory, the St\"uckelberg function has to satisfy the constraint \cite{Kodama}:

\begin{equation}
(T_0'(r,t))^2=\frac{1-f(Sr)}{f(Sr)}\left(\frac{S^2}{f(Sr)}-\cdot{T}_0^2\right).
\end{equation}
A global solution of this previous constraint, is given by the Finkelstein-type form \cite{Kodama}:

\begin{equation}   \label{eq:Tzero}
T_0(r,t)=St\pm\int^{Sr}\left(\frac{1}{f(u)}-1\right)du.
\end{equation}
Another equivalent solutions, can be found in \cite{gabagabaga}, where the extra-degrees of freedom will appear inside the fiducial metric. The present approach translates all the degrees of freedom to the dynamical metric.  

\section{The possible paths for a particle coming from the black-hole}   \label{eq:pathspart}

\begin{figure}
	\centering
	\includegraphics[width=15cm, height=8cm]{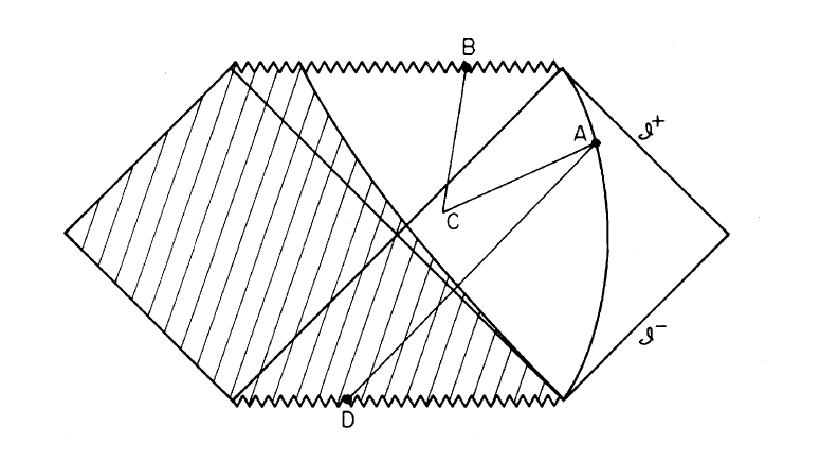}
	\caption{The Penrose diagram for the analytical extended Schwarzschild solution. Fig. Taken from \cite{Hartle}.}
	\label{fig:1}
\end{figure}

The path integral method in GR, was introduced by Hartle and Hawking in \cite{Hartle}. Here I will use the same method in order to derive the results of the Hawking radiation in massive gravity and then analyze the possible contributions coming from the extra-degrees of freedom. In the Feynman path integral method, the amplitude $K(x,x')$ for a particle to propagate from a point $x$ to another point $x'$, is given by:

\begin{equation}
K(x,x')\backsim \Sigma_{paths}e^{iS(x,x')/\hbar},
\end{equation}
where $S(x,x')$ is just the classical action for a particular path connecting $x$ and $x'$. The amplitude $K(x,x')$ is called propagator. In Fig. (\ref{fig:1}), we can observe the analytically extended Schwarzschild space. 
The shaded part of the diagram should be replaced by a gravitationally collapsing object. The path $BCA$, represents the process of a pair of particles being created at the point $C$. One of the particles propagates forward in time and reach the point $A$, where an observer with a detector is located. The other particle propagate backwards in time, reaching eventually the future singularity at $B$. There are two kind of observers in dRGT massive gravity. The first one defines the time in agreement with the St\"uckelberg function $T_0(r,t)$. This observer will not see any difference between GR and the dRGT formulation of massive gravity. This observer cannot detect any contribution coming from the extra-degrees of freedom of the theory. The second kind of observer, defines the time arbitrarily in agreement with $t$. In general, the directions of the Killing vectors defined in agreement with $T_0(r,t)$ and $t$, will disagree. When this happens, we have the possibility of extra-particle creation process due to the vacuum degeneracy generated by the extra-degrees of freedom of the theory. In general metric solutions, the extra-particle creation process will emerge from the time-dependence ($t$) of the St\"uckelberg function. In the stationary case with $A(r)$ being time-independent, which is the one explored in this manuscript, the St\"uckelberg function is linear with respect to $t$ and there are two effects that will appear still at this level. The first one is the branch point condition reproduced by the factor $S$ appearing in the metric components (\ref{eq:drgt metric}). The second effect is the shift of the poles of the propagator with respect to the GR case, even if $S=1$. Later in this manuscript, I will explain in detail both effects. For the moment, I will define the path integral method and I will explain how can it be applied to the dRGT massive gravity case. Here I do not consider the paths coming from $\mathscr{I}^-$, because they correspond to propagation of incoming particles from the infinite past. Here I will assume that the infinite past does not contain any degree of freedom. In such a case, then there is no vacuum degeneracy at that level.  
The paths corresponding to the shaded region are ignored, since they correspond to the collapsing object itself. The propagation is then considered from the future singularity. If we want to use the stationary phase method in order to derive the particle creation process, we immediately find that it is impossible to find real stationary paths connecting the future singularity with a positive-frequency mode for a stationary exterior observer \cite{Hartle}. Then we have to analytically extend the path toward the past singularity, or equivalently, we have to translate the point $B$ on the future singularity to complex values. This is equivalent to relate the rate of emission to the rate of absorption of the black-hole after thermodynamic equilibrium is reached. Mathematically, the connection is \cite{Hartle}:

\begin{equation}   \label{eq:Notation}
N(E)=P(E)e^{-2\pi E/\kappa},
\end{equation}
where $N(E)$ and $P(E)$ correspond to the probability emission and absorption respectively. Here $\kappa$ is the surface gravity of the black-hole, which in the standard case of GR becomes $\kappa=1/4GM$. In massive gravity, some modification is expected for observers located at scales larger than $r_V$ if they define their corresponding time-like Killing vectors with respect to the usual notion of time coordinate. If we want to apply the path integral method for the analysis of black-holes in GR to the dRGT case, the St\"uckelberg function plays the role of the time-coordinate as has been explained in \cite{On the app}. Then the time-coordinate $t$, will appear implicitly inside this function. This is equivalent to use the St\"uckelberg trick at the non-linear level explained in the coming section.

\section{The non-linear St\"uckelberg trick}   \label{eq:non-linearStu}

The coming section will require to use the St\"uckelberg trick in order to reproduce the important results related to the Hawking radiation inside the non-linear formulation of massive gravity. At the non-linear level, the trick looks like the dipheomorphism transformations in GR. In unitary gauge (with a fiducial metric being Minkowski), it can be expressed as \cite{K}:

\begin{equation}   \label{eq:gmunu}
g_{\mu\nu}\to G_{\mu\nu}=\frac{\partial Y^\alpha}{\partial x^\mu}\frac{\partial Y^\beta}{\partial x^\nu}g_{\alpha\beta}(Y(x)),
\end{equation}
where $Y^\alpha$ are the components of the St\"uckelberg functions. The previous equation looks like a standard gauge transformation in GR. It is however, the way of introducing redundant variables in order to restore the dipheomorphism invariance of the massive action in massive gravity. The redundant variables contain the information of the extra-gravitational degrees of freedom. Under the dipheomorphism transformations given by:

\begin{equation}   \label{eq:gt}
g_{\mu\nu}\to\frac{\partial f^\alpha}{\partial x^\mu}\frac{\partial f^\beta}{\partial x^\nu}g_{\alpha\beta}(f(x)), \;\;\;\;\;Y^\mu(x)\to f^{-1}(Y(x))^\mu,
\end{equation}
the metric (\ref{eq:gmunu}) is invariant. This is necessary for keeping the massive action invariant. Infinitesimally, the St\"uckelberg function expansion:

\begin{equation}   \label{eq:Thisy}
Y^\alpha(x)= x^\alpha+A^\alpha(x),
\end{equation}
provides the following result on the metric (\ref{eq:gmunu}):

\begin{eqnarray}   \label{eq:ob}
G_{\mu\nu}\approx g_{\mu\nu}+A^\lambda\partial_\lambda g_{\mu\nu}+\partial_\mu A^\alpha g_{\alpha\nu}+\partial_\nu A^\alpha g_{\alpha\mu}+\frac{1}{2}A^\alpha A^\beta\partial_\alpha\partial_\beta g_{\mu\nu}+\partial_\mu A^\alpha\partial_\nu A^\beta g_{\alpha\beta}\nonumber\\
+\partial_\mu A^\alpha A^\beta\partial_\beta g_{\alpha\nu}+\partial_\nu A^\alpha A^\beta\partial_\beta g_{\mu\alpha}+...
\end{eqnarray}
The gauge transformations (\ref{eq:gt}), after the infinitesimal expansion $f(x)=x+\zeta(x)$, become:

\begin{equation}
\delta g_{\mu\nu}=\zeta^\lambda\partial_\lambda g_{\mu\nu}+\partial_\mu\zeta^\lambda g_{\lambda\nu}+\partial\zeta^\lambda g_{\mu\lambda},
\end{equation}

\begin{equation}
\delta Y(x)= -\zeta^\mu(Y),\;\;\;\;\;\delta A^\mu=-\zeta^\mu-A^\alpha\partial_\alpha\zeta^\mu-\frac{1}{2}A^\alpha A^\beta\partial_\alpha\partial_\beta\zeta^\mu-...
\end{equation}
The $A^\mu$-term corresponds to the Goldstone bosons that at the non-linear level carry the broken symmetry in massive gravity. It can be verified again that under the previous infinitesimal gauge transformations, eq. (\ref{eq:ob}) provides the result:

\begin{equation}
\delta G_{\mu\nu}=0,
\end{equation}
which demonstrates that the dynamical metric with the St\"uckelberg functions appearing explicitly is dipheomorphism invariant. 

\section{Analyticity properties of the propagator}   \label{eq:Analitycity}
If we want to derive the analyticity properties of the propagator, the easiest way is by using the St\"uckelberg trick, already explained in the previous section. The trick permits us to define the St\"uckelberg function in terms of new objects which can be considered as extended coordinates. The basic idea is to replace the standard time-coordinate $t$ by the use of the St\"uckelberg function $T_0(r,t)$ inside this new objects. We can then define the components of the St\"uckelberg functions as extended coordinates. Here we start with the basic differential definition of the propagator given by:

\begin{equation}   \label{eq:waveequation}
(\square^2-m^2)K(x,x')=-\delta(x,x'),
\end{equation}
with the appropriate boundary conditions. In dRGT massive gravity, the condition $\beta=(3/4)\alpha^2$ is equivalent to the Schwarzschild solution (asymptotically flat), but with some background of St\"uckelberg fields. Inside the Schwarzschild geometry, we will consider the case where $x'$ is external to the black-hole and $x$ is over the horizon as in the standard case. If the Vainshtein mechanism operates, as it is expected from these solutions, then the extra-degrees of freedom effects will become relevant after the Vainshtein scale and then the time coordinate orientation selected by the observer becomes important for the vacuum definition and as a consequence, for the definition of temperature. As has been explained before, for observers defining the time in agreement with $T_0(r,t)$, the perceived radiation will not change with respect to the GR case \cite{1, On the app}. On the other hand, for observers defining the time in the standard way, the effect of the extra-degrees of freedom will appear and they will generate changes in the periodicity pattern for the poles of the propagator. We can define the Kruskal coordinates in dRGT by using the St\"uckelberg trick. They are defined as \cite{1, On the app}:

\begin{equation}   \label{eq:condo1}
ds^2=-\left(\frac{32M^3e^{-r/2GM}}{r}\right)dU'dV'+r^2d\Omega^2,
\end{equation}
with:

\begin{equation}   \label{eq:condo}
U'V'=\left(1-\frac{r}{2GM}\right)e^{r/2GM},
\end{equation}
where the St\"uckelberg fields components are defined as:

\begin{equation}   \label{eq:adv}
V'=\left(\frac{r}{2GM}-1\right)^{1/2}e^{\left(r+T_0(r,t)\right)/4GM},
\end{equation}

\begin{equation}   \label{eq:adv35678}
U'=-\left(\frac{r}{2GM}-1\right)^{1/2}e^{\left(r-T_0(r,t)\right)/4GM},
\end{equation}
and $T_0(r,t)$ is the initial St\"uckelberg function. The objects (\ref{eq:adv}) and (\ref{eq:adv35678}) are the new St\"uckelberg functions. They can also be defined as extended coordinates inside the dRGT formulation. Note that the condition (\ref{eq:condo}) is not affected by the presence of the extra-degrees of freedom of the theory (dRGT) because the initial St\"uckelberg function ($T_0(r,t)$) does not appear in the final result. In fact, whenever we find time-independent quantities defined originally inside the framework of GR, when extended to the dRGT case, will not be affected by the presence of the extra-degrees of freedom in spherically symmetric configurations. Only time-dependent quantities as they are defined originally inside the framework of GR, will suffer some changes when they are extended to the dRGT case. The definition of event horizon taken from the combination of St\"uckelberg function components in eq. (\ref{eq:condo}) is unchanged with respect to the GR case. Then the event horizon is defined with the condition $U'=0$ or $V'=0$, when $r=2GM$. This can also be verified from the metric (\ref{eq:drgt metric}), then the event-horizon position will not be affected by the presence of the extra-degrees of freedom. The complexified horizon in this case, is defined as the surface where the non-zero of the coordinates $U'$ or $V'$ is extended analytically to complex values. Any singularities for the propagator will come from the point $W=0$ as in the standard case \cite{Hartle}. By dividing the interval of $W$ as in \cite{Hartle}, the propagator becomes:

\begin{equation}
K(x,x')=K_0(x,x')-i\Sigma_c\frac{e^{is_c(x,x')/4W_0}}{s_c(x,x')+i\epsilon}D_c(x,x'),
\end{equation}
where $K_0(x,x')$ corresponds to the propagator of the interval $[0,W_0]$ for the parameter $W$ with $W_0$ being infinitesimal and near the divergence point of the propagator. We can observe that there is a singularity as $s_c(x,x')=-i\epsilon$. In massive gravity, the extra-degrees of freedom are contained inside this condition if we take into account the St\"uckelberg trick as has been formulated in eq. (\ref{eq:gmunu}). Then in terms of the St\"uckelberg function, this result corresponds to a null geodesic, connecting $x'$ with the complexified manifold. When expressed explicitly in terms of the standard time coordinate $t$, this previous null condition contains some extra-terms (St\"uckelberg fields) which will modify the analyticity properties of the propagator. All the geodesics starting from real values of $x'$, will intersect the horizon at real sections, namely, sections where the extended versions of $V'$ and $U'$ are real. We can consider the geodesics which connect a real value of $x'$ which is outside the black-hole with the future event horizon. Then we can define again new St\"uckelberg functions by extending the notion of the advanced Eddington-Finkelstein coordinates in agreement with (\ref{eq:adv}), covering $V'\geq0$:

\begin{equation}   \label{eq:this one}
V'=e^{\kappa V},
\end{equation}
with $V$ representing the extended version (St\"uckelberg function) of the advanced-null coordinates $V=v+A(r,v)$. Here $T_0(r,t)=St+A(r,v)$. At this point of the calculation, we have two options. The first one is to work by using $V$ as the Killing time. In such a case, we will obtain exactly the same results of GR. This corresponds to the observers defining the vacuum in agreement with $T_0(r,t)$. The second option, is to use $v=V-A(r,t)$ as the Killing time, defining then eq. (\ref{eq:this one}) as:

\begin{equation}
V'=e^{\kappa (v+A(r,v))}.
\end{equation}
It is possible for some observers to define the vacuum in agreement with $v$, namely, the usual time-coordinate. By complexifying the geodesics assuming an imaginary affine parameter $\lambda$, then the geodesic would be a two-dimensional sheet in the complex coordinate system and then from eq. (\ref{eq:this one}), we know that for complex extensions of $V$, there is a periodicity associated to the coordinate $V'$ and given by $Im\kappa=2\pi/V$ or:

\begin{equation}
Im\kappa=\frac{2\pi}{(v+A(r,v))}.
\end{equation}
Note however that if we use the coordinate $v$ as the Killing time, then any dependence on $v$ coming from the function $A(r,v)$, will affect the periodicity properties of the complexified variable $v$. For stationary solutions with $A(r)$ being independent of $v$, like the one explored in this manuscript, the periodicity structure of the propagator still can be affected by a couple of effects that can appear due to the presence of extra-degrees of freedom. These effects will be explained later in detail. What is important for the moment is to note that the periodicity associated with the variable $V$, can be different to the one associated to the variable $v$. By convenience, we select the strip which contains real values of $V'$. We suppose that for some affine parameter $\lambda=0$, the coordinates (and St\"uckelberg functions) take the values $V'_0,r_0,\theta_0,\phi_0$, which are assumed to be real. Then we can study which complex values of the extended coordinate (St\"uckelberg function) $V$ in the strip with real values of $r=2GM$, $\theta$ and $\phi$ are contained inside the two-dimensional sheet generated by the null geodesics. It is clear that the two-dimensional sheet generated by the complex values of $V$, when expressed in terms of $v$, will in general contain some arbitrary combinations of the complex values of $v$ and $A(r,v)$, depending on the specific form of the St\"uckelberg function $T_0(r,v)$. 
By extending the method proposed by Hartle and Hawking \cite{Hartle}, the singularities of the propagator on the complexified horizon are slightly displaced from the real values of U' and V'. They correspond to the poles marked by $s(x,x')=-i\epsilon$, which are infinitesimal displacements with respect to the real values on the complexified horizon. The direction of the displacements, can be found in the same way as Hartle and Hawking did \cite{Hartle}. The analyticity properties of the propagator suggests that the periodicity of the poles will change with respect to the GR case for observers taking the time in agreement with $t$ defined in an arbitrary direction with respect to $T_0(r,t)$. Observers defining the time in agreement with $T_0(r,t)$, will not perceive any change in the periodicity of the poles of the propagator. For these special observers, there is no difference with respect to the GR case. We can define St\"uckelberg functions in agreement with the extended null-coordinates $U'$ and $V'$ defined as:

\begin{eqnarray}   \label{eq:Munsang}
U'=\left(1-\frac{r}{2GM}\right)^{1/2}e^{(r-T_0(r,t))/4GM},\nonumber\\
V'=\left(1-\frac{r}{2GM}\right)^{1/2}e^{(r+T_0(r,t))/4GM},
\end{eqnarray}
with the (initial) St\"uckelberg function $T_0(r,t)$ playing the role of an extended notion of time. The results (\ref{eq:Munsang}) are valid for the cases $U'>0$ and $V'>0$. On the other hand, we also have:

\begin{eqnarray}   \label{eq:Munsang2}
U'=-\left(\frac{r}{2GM}-1\right)^{1/2}e^{(r-T_0(r,t))/4GM}, \nonumber\\
V'=\left(\frac{r}{2GM}-1\right)^{1/2}e^{(r+T_0(r,t))/4GM},
\end{eqnarray}
for the cases $U'<0$ and $V'>0$. Analogous relations can be found for the other quadrants of the extended Penrose diagram. 
\begin{figure}
	\centering
		\includegraphics[width=15cm, height=8cm]{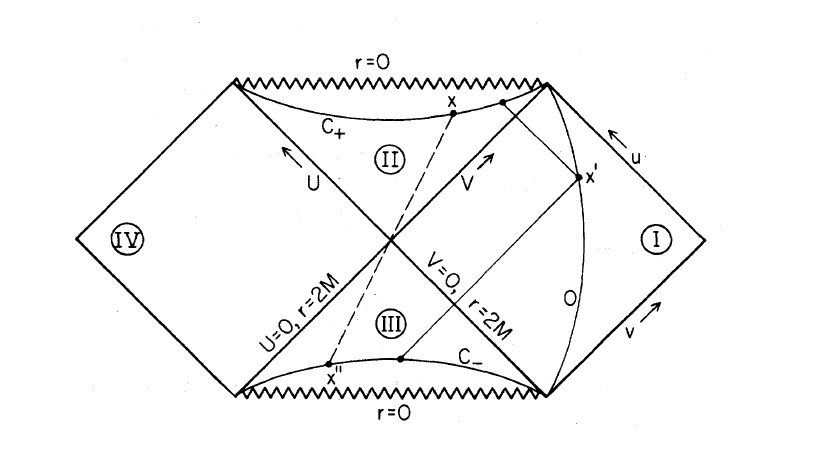}
	\caption{The Penrose diagram for the Schwarzschild geometry. In dRGT this diagram would correspond to the one described by observers defining the time in agreement with the St\"uckelberg function $T_0(r,t)$. Taken from \cite{Hartle}.}
	\label{fig:momoko}
\end{figure}
In the diagram (\ref{fig:momoko}), the region $II$ has the Cauchy data over part of the future horizon and part of the past horizon. Then the initial data over these surfaces, determine uniquely the propagator inside this region. If we complexify the original St\"uckelberg function $T_0(r,t)$, and keeping the coordinates $r,\theta,\phi$ real, then eqns. (\ref{eq:Munsang}) can be written in the modulus-argument form:

\begin{equation}
U'=\vert U'\vert e^{-i\psi(r,t)/4GM}, \;\;\;\;\;\;\; V'=\vert V'\vert e^{i\psi(r,t)/4GM},
\end{equation}
where $T_0(r,t)=\gamma(r,t)+i\psi(r,t)$. Then the problem of determining the propagator is reduced to solve the wave equation (\ref{eq:waveequation}), in the real functions $\vert U'\vert$ and $\vert V'\vert$ for fixed values of $\psi(r,t)$. This is analogous to the standard case illustrated by Hawking and Hartle \cite{Hartle}. However, the main difference here is the fact that the function $\psi(r,t)$, contains the information of the extra-degrees of freedom of the theory. The propagator is analytic on the complexified horizon on the upper-half plane of the complex variable $U'$ and on the lower half plane for the complex variable $V'$. The Cauchy data is regular when:

\begin{equation}   \label{eq:mylove}
-4\pi GM<\psi(r,t)<0,
\end{equation}
where $\psi(r,t)=\mu+\bar{A}(r,t)$, with $\bar{A}(r,t)$ being the imaginary component of $A(r,t)$. If we know the exact dependence of the function $A(r,t)$, then we can separate easily the real part from the imaginary one. The extra-component of radiation, coming from the deformation of the periodicity pattern of the propagator produced by the St\"uckelberg function $T_0(r,t)$, depends on whether or not there is an imaginary component for $A(r,t)$ after doing the analytical extension on the variable $t$. The observers defining the time in agreement with $T_0(r,t)$, will never be able to perceive any extra-component of radiation and they will define the same result of temperature as in GR. On the other hand, observers defining the time in an arbitrary direction with respect to $T_0(r,t)$, might be able to perceive an extra-component of radiation, depending on the explicit form of the St\"uckelberg function. The propagator is analytic in $T_0(r,t)$ in the same form as it appears in GR. The functions $\bar{U'}$ and $\bar{V'}$ satisfy the Cauchy-Riemann (orthogonality) conditions, already demonstrated in \cite{Hartle}. Then given the previous arguments, the propagator $K(x,x')$ will be analytic in $T_0(r,t)$ in a strip of width $4\pi M$. Notice however, that this means that:

\begin{equation}   \label{eq:regularity}
-4\pi GM<\mu+\bar{A}(r,t)<0.
\end{equation}
If we want to solve for $\mu$ this previous inequality, we have to take into account that $A(r,t)$ can be time-dependent. This means that this function can be written as:

\begin{equation}
A(r,t)=A(r,\gamma+i\mu),
\end{equation}
where we have defined $t=\gamma+i\mu$. The exact range of regularity for the propagator given by eq. (\ref{eq:regularity}), and solved in terms of the standard complex time coordinate will depend on the exact form of the function $A(r,t)$. Only if this function is time-independent, then it is possible to solve trivially for $\mu$ in eq. (\ref{eq:regularity}) as follows:

\begin{equation}   \label{eq:compexala}
-4\pi GM-\bar{A}(r)<\mu<-\bar{A}(r).
\end{equation}
From these previous results, we can observe that the periodicity of the poles of the propagator might change due to the presence of the extra-degrees of freedom. In other words, the periodicity associated to the variable $t$, is not necessarily the same periodicity associated to the function $T_0(r,t)$. The exact behavior of the poles of the propagator with respect to any observer, depends on the exact functional dependence of the St\"uckelberg function $T_0(r,t)$.
The modified period of the propagator, will reproduce a change on the amount of particles detected if the detectors operate by using the usual notion of time $t$. The Vainshtein radius marks the scale after which the extra-degrees of freedom become relevant, then it is expected that the effects explained in this manuscript are relevant after the Vainshtein radius \cite{1}. If we use the St\"uckelberg function as the time-coordinate, then the Penrose diagram for the case with zero cosmological constant will not differ from the standard Schwarzschild diagram shown in Fig. (\ref{fig:1}) \cite{1}. Then the propagator $K(x,x')$ with $x$ in the region $II$ of the diagram (\ref{fig:momoko}), still can be continued to $T_0(r,t)\to T_0(r,t)-i4\pi M$. This analytical extension, simply relate the propagator connecting the points $x$ and $x'$, with the one which connects the points $x'$ and $x''$ in the diagram (\ref{fig:momoko}). Note that $x''$ is in the region $III$. The Penrose diagram (\ref{fig:momoko}), contains implicitly the extra-degrees of freedom of the theory. If the observers define the time in agreement with $T_0(r,t)$, then everything will look like GR. These special kind of observers will never perceive any difference between massive gravity and GR and they will describe the physics in agreement with the standard Penrose diagram. 
The analytical extension just mentioned, marks the relation between the rates of emission and absorption of the black-hole. If we want to understand the physical effects of the extra-degrees of freedom, then we have to express the propagator in terms of the ordinary time $t$. Then for detectors operating with respect to $t$, instead of $T_0(r,t)$, the analytic extension does not necessarily relates the rate of emission with the rate of absorption of the black-hole radiation. This is because the extension is related not only to the complex part of the ordinary time variable $t$, but also to the complex part of the function $A(r,t)$, namely $\bar{A}(r,t)$. 
The analyticity of the propagator is periodic as in the standard case with period $8\pi GM$ for observers describing the physics with respect to $T_0(r,t)$ \cite{Hartle}. For the case of observers describing the physics with respect to the usual time $t$, it is not clear at all what is the periodicity associated to the complex time. Later I will demonstrate that for the case under study in this manuscript, the periodicity structure of the propagator, depends on the values of the free-parameters of the theory.

\subsection{The analyticity of the propagator: The case $\beta=(3/4)\alpha^2$}   \label{eq:miaush}
\begin{figure}
	\centering
		\includegraphics[width=12cm, height=8cm]{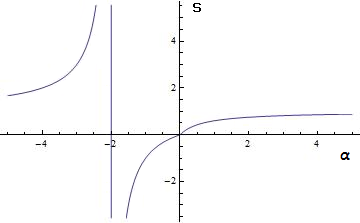}
	\caption{The parameter $S$ and a function of $\alpha$ for the case $\beta=(3/4)\alpha^2$. Note that the function has a discontinuity for the case $\alpha=-2$.}
	\label{fig:momokomiaumiau}
\end{figure}

The explicit St\"uckelberg function $T_0(r,t)$, can be found for the case $\beta=(3/4)\alpha^2$ from eq. (\ref{eq:Tzero}), the result after integration is:

\begin{equation}   \label{eq:tsubzerola2}
T_0(r,t)=St\pm\frac{2GM}{S}Log\left\vert1-\frac{rS}{2GM}\right\vert.
\end{equation} 
This result looks similar to the tortoise coordinates introduced in GR. The argument of the natural Logarithm has a different behavior in this case however. The scale at which the logarithm diverges, depends on the exact value of $S$. The scale factor $S$ is a function of the free-parameters of the theory \cite{Kodama}. If we compare the result (\ref{eq:tsubzerola2}) with the definition of tortoise coordinate given by:

\begin{equation}   \label{eq:rtortoise}
r^*=r\pm2GMLog\left\vert\frac{r}{2GM}-1\right\vert,
\end{equation}
we immediately notice the similarity in both cases. Note that if $r=0$ (The singularity point), the expression (\ref{eq:tsubzerola2}) becomes $T_0(r,t)=St$, this means that when the gravitational field becomes stronger, the extra-degrees of freedom become negligible as it should be if the Vainshtein mechanism operates. Then at scales close to the source, GR is approximately recovered. The scale factor $S$, as a solution of the parameters of the theory and for the case $\beta=(3/4)\alpha^2$ can be expressed as:

\begin{equation}   \label{eq:Theparamerela}
S=\frac{3\alpha}{4+3\alpha-\frac{2\alpha}{\vert\alpha\vert}}. 
\end{equation}     
Note that this function becomes equivalent to:

\begin{eqnarray}
S=\frac{3\alpha}{4+3\alpha+2},\;\;\;if\;\;\;\alpha<0,\;\;\;\;\;S=\frac{3\alpha}{4+3\alpha-2},\;\;\;if\;\;\;\alpha>0. 
\end{eqnarray} 
Is interesting to observe that as $\alpha\to\infty$ or $\alpha\to-\infty$, then $S\to1$. Note that $S$ is a well behaved function, except when $\alpha=-2$ which represents a point of discontinuity. The figure (\ref{fig:momokomiaumiau}) illustrate the behavior of $S$ as a function of $\alpha$. 
If we select the parameter $\alpha$ to be very small, then one of the two possible branches of $S$ (see fig. (\ref{fig:momokomiaumiau})) is also small and the distance where the St\"uckelberg function (\ref{eq:tsubzerola2}) diverges can be very large. The scale of divergence is given by:

\begin{equation}
r_H=\frac{2GM}{S},
\end{equation} 
which is related to the scale of the black-hole. When we analyze the Hawking radiation, usually we use the advanced or retarded coordinates in GR. In massive gravity, we can define in analogous way, the St\"uckelberg functions as extended advanced and retarded coordinates in agreement with:

\begin{equation}   \label{eq:extenddadv}
V=T_0(r,t)+Sr+\frac{2GM}{S}log\left\vert\frac{Sr}{2GM}-1\right\vert,
\end{equation}

\begin{equation}   \label{eq:extended ret}
U=T_0(r,t)-Sr-\frac{2GM}{S}log\left\vert\frac{Sr}{2GM}-1\right\vert,
\end{equation}
where the re-scaling factor $S$ has been used. Here there are two differences with respect to the GR case; the first one is the replacement of the time coordinate by the St\"uckelberg function $T_0(r,t)$. The second one is the inclusion of the factor $S$ re-scaling the distance coordinate $r$. We then get different results depending on the combination of signs between $T_0(r,t)$ and $r^*$ defined in agreement with eq. (\ref{eq:rtortoise}) but taking into account the re-scaling factor $S$. The following are the possible cases obtained when we extend the notions of advanced and retarded coordinates in order to define the St\"uckelberg functions:

\subsection{Case I:}

\begin{equation}   \label{eq:extenddadvaa}
V=S(t+r)+2\pi i\frac{GM}{S}+\frac{4GM}{S}log\left\vert\frac{Sr}{2GM}-1\right\vert,
\end{equation}

\begin{equation}   \label{eq:extended retaa}
U=S(t-r)+2\pi i\frac{GM}{S}.
\end{equation}

\subsection{Case II:}

\begin{equation}   \label{eq:extenddadvaaa}
V=S(t+r)-2\pi i\frac{GM}{S},
\end{equation}

\begin{equation}   \label{eq:extended retaaa}
U=S(t-r)-2\pi i\frac{GM}{S}-\frac{4GM}{S}log\left\vert\frac{Sr}{2GM}-1\right\vert,
\end{equation}
and combinations of these previous results. The logarithmic argument in eqns. (\ref{eq:extenddadvaa}) and (\ref{eq:extended retaaa}) is positive defined when:

\begin{equation}
r\geq \frac{2GM}{S}.
\end{equation}
Otherwise the Logarithmic function provides a complex result. We can do a new re-definition of the St\"uckelberg function as follows:

\begin{equation}
V'=e^{SV/4GM},\;\;\;\;\;U'=-e^{-SU/4GM},
\end{equation}
taking into accont the implicit appearance of the extra-degrees of freedom, then from eqns. (\ref{eq:extenddadvaa}), (\ref{eq:extended retaa}), (\ref{eq:extenddadvaaa}) and (\ref{eq:extended retaaa}), we get:

\subsection{Case I:}

\begin{equation}
V'=i\left(\frac{Sr}{2GM}-1\right)e^{S^2(t+r)/4GM},
\end{equation}

\begin{equation}
U'=ie^{S^2(r-t)/4GM},
\end{equation}

\begin{figure}
	\centering
		\includegraphics[width=14cm, height=7cm]{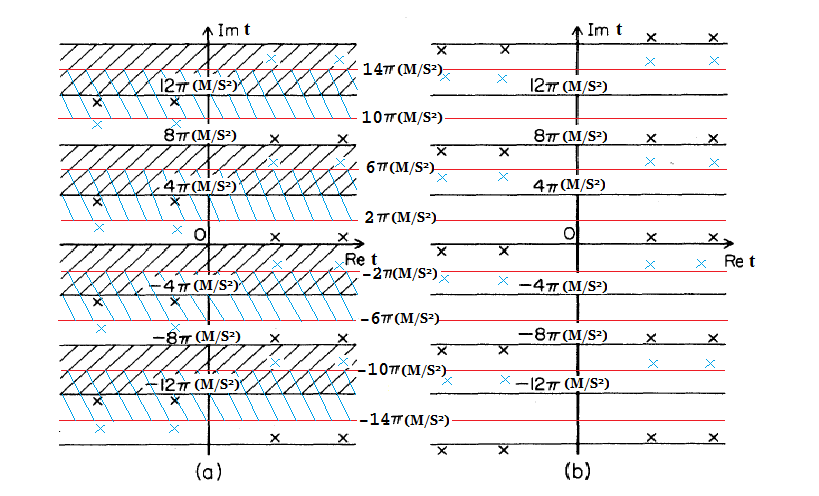}
	\caption{Modified periodicity structure of the propagator due to the presence of the extra-degrees of freedom. The red lines and blue regions represent the analyticity regions in agreement with an observer defining the time in agreement with $t$. The black lines and regions represent the analyticity regions perceived by the observers defining the time in agreement with the St\"uckelberg function $T_0(r,t)$ if $S=1$.}
	\label{fig:alejo}
\end{figure}

\subsection{Case II:}

\begin{equation}
V'=ie^{S^2(t+r)/4GM},
\end{equation}

\begin{equation}
U'=-i\left(\frac{Sr}{2GM}-1\right)e^{S^2(r-t)/4GM},
\end{equation}
and combination of these cases. If we analytically extend the time-coordinate to complex values, then we have $t=\tau+i\mu$. And then the St\"uckelberg functions defined in agreement with $V'$ and $U'$ can be written in the form:

\begin{equation}
V'=\vert V'\vert e^{i(S^2\mu/4GM+\pi/2)},\;\;\;\;\;U'=\vert U'\vert e^{-i(S^2\mu/4GM+\pi/2)},
\end{equation}
for any of the cases illustrated before. The Cauchy data is then regular when the condition:

\begin{equation}
-\frac{6\pi GM}{S^2}<\mu<-\frac{2\pi GM}{S^2},
\end{equation}
is satisfied. If we compare this result with the notation used in eq. (\ref{eq:compexala}), then the periodicity of the propagator is related to the complex function:

\begin{equation}
\bar{A}(r,t)=\frac{2\pi GM}{S^2}.
\end{equation}
Then for the case $\beta=(3/4)\alpha^2$, the observers defining the physics in agreement with $t$, will perceive two effects, the first one is the shift of the period with respect to GR, even if $S=1$. In this case, the periodicity associated to the propagator can be the same as in GR, but the shift of the positions of the poles, create the effects of time-delay for the detection of particles. In other words, if an observer inside the theory of GR has a detector at scales located after $r_V$, he will disagree with the instant of times at which the particles appear with respect to the case of an observer located at the same distant and working inside the dRGT formulation of massive gravity. However, both observers will agree in the average temperature obtained. The second effect inside dRGT is produced by the factor $S$. This factor opens the possibility of having branch points if it takes a non-integer value. In fact, $S^2$ represents the order of the branch point with respect to the GR case. Then the order of the branch point will depend on the parameters of the theory. For the case explored in this manuscript, it will depend on $\alpha$. Note that for $\alpha=-2$, the branch point is of zero order and the detection of particles is suppressed. If $\alpha=\pm\infty$, then the branch point condition disappears and the theory looks as GR except for the fact of the disagreement of the instant at which the particles appear when we compare the GR case with respect to the dRGT one. Finally, if $\alpha=0$, the branch point is of infinite order one and the particle creation process is also suppressed. Fig. (\ref{fig:alejo}) illustrate the behavior of the of periodicity structure of the propagator for observers defining the time in agreement with $t$, with respect to observers defining the time in agreement with $T_0(r,t)$. 

\section{Black hole radiance}   \label{eq:Blackrad}

In the GR case, when we analyze the amplitude of emission of a mode with definite positive energy $E$. Its time-dependence is given by $f\backsim exp(-iEt)$. The information about emission is contained in the amplitude $\varepsilon_E$, which is defined as \cite{Hartle}:

\begin{equation}   \label{eq:Momokocheung.com}
\varepsilon_E(\vec{R}', \vec{R})=\int^{+\infty}_{-\infty}dT_0(r,t)e^{-iET_0(r,t)}K(T_0(r,t),\vec{R};0,\vec{R}').
\end{equation}
In this case however, the ordinary time coordinate has been replaced by the St\"uckelberg function $T_0(r,t)$. Here, $\varepsilon_E$ is the component of energy, coming in principle from the surface $C_+$ as in the GR case. However, in massive gravity the effect of the extra-degrees of freedom will appear through the St\"uckelberg function $T_0(r,t)$. Inside the GR formulation, $C_+$ is the only surface which contributes to the radiation perceived by the detector located at large scales. In massive gravity however, since the extra-degrees of freedom appear, it is possible that some other surfaces can contribute. Here however, we can still assume that the pre-collapse surface does not contribute to the final result. This means that for a detector confined to work inside an interval $(-t'_1,t'_1)$, with $t_1'$ very large, still we can assume the absence of particles in the infinite past. In massive gravity however, we would be talking about an interval defined by the St\"uckelberg function. Then for some observers, the pre-collapse surface given at the time $T_0(r,t)=t'_1\to-\infty$, would be equivalent to $St'+A(r,t)\to-\infty$. Then the assumption of a vacuum devoid of particles in the space-like pre-collapse surface for any observer, is equivalent to say that the extra-degrees of freedom are also absent there. Even if still the contributions due to the extra-degrees of freedom will not appear from the pre-collapsing surface, they might appear from the time-like surface connecting the space-like surface inside the future horizon with the observation surface, where the detectors are located. When the detectors are located at scales larger than $r_V$, the contribution coming from the extra-degrees of freedom becomes relevant. For an observer located at scales larger than $r_V$ and defining the time arbitrarily with respect to $T_0(r,t)$, the perceived effect is equivalent to some extra-modes not necessarily coming from the event horizon of the black-hole. 
In the spherically symmetric solution in massive gravity found in \cite{Kodama}, the background solution has a symmetry under time-translations. The quantity conserved under time-translations is however a combination of the usual notion of energy inside GR, with a velocity-dependent quantity associated to the extra-degrees of freedom \cite{On the app, Momo, My paper 1}. This conserved quantity defines the notion of energy in agreement with the observers defining the time with respect to the St\"uckelberg function. Then the propagator will be a function of $T'_0(r,t)-T_0(r,t)$ at large values of $T_0(r,t)$. Observers defining the time in agreement with $T_0(r,t)$, will not perceive any contribution coming from the time-like surface connecting the space-like surface inside the future event horizon, with the space-like surface of observation. On the other hand, the observers defining the time arbitrarily, will perceive some (possible) extra-contribution. The extra-contribution will appear as a change in the periodicity of the propagator. The extra-component of radiation, will become relevant at scales larger than $r_V$. Now we can relate the rate of emission with the rate of absorption. For the detectors defining the time in agreement with $T_0(r,t)$, the result is trivial and it is the same to the one obtained in GR \cite{Hartle}. In this trivial situation, by distorting the contours of integration in $T_0\to T_0-4\pi Mi$, then eq. (\ref{eq:Momokocheung.com}), becomes:

\begin{equation}   \label{eq:Eva}
\varepsilon_E(\vec{R}',\vec{R})=e^{-4\pi ME}\int^{+\infty}_{-\infty}dT_0e^{-iET_0(r,t)}K\left(T_0(r,t)-4\pi Mi,\vec{R};0,\vec{R}'\right).
\end{equation}
If we however expand eq. (\ref{eq:Eva}) by using the usual notion of time and separating real and imaginary contributions of the St\"uckelberg function in agreement with $T_0(r,t)=St+i\mu+A(r,t)+i\bar{A}(r,t)$, then there is no guarantee that the previous result relates the rate of emission with the rate of absorption for the standard Penrose diagram, namely, the one written with respect to the usual notion of time $t$. Then in such a case, the extended relation between the rates of emission and absorption, not only includes the radiation coming from the event horizon, but also some extra-component appearing at scales larger than $r_V$. This extra contribution comes from the time-like surface connecting $C_+$ with the space-like observation surface. Similar conclusions were found in \cite{1}. 

\subsection{The radiation with respect to observers defining arbitrary time}

When the degrees of freedom inside the dynamical metric appear explicitly, and the detectors are peaked in agreement with the usual time-coordinate, then some extra-contribution to the radiation process might appear at scales larger than $r_V$. In agreement with the previous analysis of the analyticity of the propagator, there are poles when $s(x,x')=-i\epsilon$. An observer defining the time in agreement with the St\"uckelberg function, will define the positions of the pole propagator in agreement with the expression \cite{Hartle}:

\begin{equation}   \label{eq:Periodmiau1}
T_0(r,t)-T_0(r,t)'=\pm\left(\vert\vec{x}-\vec{x}'-i\epsilon\vert\right).
\end{equation}   
In terms of the usual notion of time, this is equivalent to:

\begin{equation}   \label{eq:Periodmiau2}
t-t'=\frac{1}{S}\left(\pm\left(\vert\vec{x}-\vec{x}'-i\epsilon\vert\right)-(A(r,t)-A'(r,t))\right).
\end{equation}
In the case under study here, this previous condition also corresponds to the singularities of the propagator. The singularities are periodic in agreement with the analysis of the previous section. It is the periodicity of the poles what reproduces the effect of particle creation. If we imagine for a moment the function $T_0(r,t)=St+A(r,t)$, with $A(r,t)$ arbitrary (initially), then the analyticity of the propagator reveals that the period associated to the St\"uckelberg function $T_0(r,t)$ is not necessarily the same period associated to the usual time-coordinate.
This is expected by intuition if we observe that in agreement with the expressions (\ref{eq:Periodmiau1}) and (\ref{eq:Periodmiau2}), the periodicity of the St\"uckelberg function $T_0(r,t)$, has to be divided by two: 1). One part belonging to the complex time variable $\mu$ as has been defined previously. 2). The other fraction going to the complex part the function $\bar{A}(r,t)$. This phenomena will in general change the perception of temperature for the detectors operating under the usual notion of time $t$ because the relation between the emission and absorption of the black-hole radiation for the stationary situation is not trivial anymore. For the stationary case analyzed in this manuscript with the parameter combination $\beta=(3/4)\alpha^2$, the relation between the emission and absorption, can be trivially derived by using eq. (\ref{eq:Eva}). The extension in this case is trivial because of the time-independence of the function $A(r)$ contained inside the original St\"uckelberg function. The trick now is to express eq. (\ref{eq:Eva}) explicitly in terms of the usual time coordinate. In general however, the result is not so trivial if the function $A(r,t)$ is time-dependent. The reason is that the delta function coming from the time-integral and representing the conservation of energy will not appear and then there is no justification for using the simple expression (\ref{eq:Momokocheung.com}) in this general situations \cite{Hartle}. However, if $A(r,t)$ is time-independent, then still there might appear a delta-function, and then the modification to the black-hole temperature as it is perceived by the observer working with the usual notion of time $t$ is trivially found. Then we get:

\begin{equation}   \label{eq:emabs}
\varepsilon_E(\vec{R}',\vec{R})=e^{-E\tau}\int^{+\infty}_{-\infty}dT_0e^{-iEt}K\left(t-i\tau,\vec{R};0,\vec{R}'\right),
\end{equation}
where $\tau$ represents the modified period. The modification of the period with respect to the GR case will depend on the extra-degrees of freedom as has been explained previously. It is clear from eq. (\ref{eq:emabs}), that the correspondence between the rates of emission and absorption related to the variable $T_0(r,t)$, does not necessarily implies the same correspondence for the rates of emission and absorption described with respect to $t$. By using these previous arguments, the temperature perceived by a detector using as a reference the time $t$ is given by: 

\begin{equation}   \label{eq:temperature}
T=\frac{1}{2\tau},
\end{equation}
This is in agreement with the result found in \cite{On the app} where however, the modified period was expressed in terms of the complex component of the function $A(r)$. For the case explored in this manuscript, if we use the results of Sec. (\ref{eq:miaush}), the exact value of temperature is given by:

\begin{equation}
T=\frac{S^2}{8\pi M}.
\end{equation}
This value of temperature can be higher or smaller than the GR case, depending on the relation $S^2/M$, which also depends on the free-parameters of the theory. 

\section{The case of effective metrics: Ghost-free coupling to matter of the reference metric}   \label{Secla}

Recently it has been demonstrated that it is possible to couple the reference metric $f_{\mu\nu}$ to matter and still keep the ghost-free condition \cite{Newder}. In this paper I have remarked that the physics at the event horizon level inside the formulation of massive gravity (the standard version) never changes. In other words, the mismatch of periodicity between the St\"uckelberg function $T_0(r,t)$ and the ordinary time-coordinate reported previously does not mean that the radiation emitted by the event horizon of the black-hole changes. What I have reported previously in this manuscript is that at large scales from the source, the vacuum will be degenerate due to the presence of the extra-degrees of freedom. This means that different observers defining different notions of time with respect to $T_0(r,t)$, will in general disagree on the amount of particles detected. This is the case because they define different notions of vacuum. We have to take into account that in general the concept of particle is not well defined inside GR due to the space-time curvature. In the same sense, in dRGT massive gravity (standard version), the concept of particle is also distorted due to the effect of the extra-degrees of freedom on the local definition of time. Then the effects described previously in this paper are relevant after the Vainshtein scale or whenever the extra-degrees of freedom are relevant. On the other hand, for the case of direct coupling to matter as it is the case explored in \cite{Newder}, it is evident that the radiation emitted by the event horizon will change because an effective metric can define a completely different scale for the horizon. Here I will only show naively the procedure to follow for this special case. In agreement with \cite{Newder}, the effective metric, able to reproduce a ghost-free theory is defined as

\begin{equation}      
g_{\mu\nu}^{eff}=\alpha^2g_{\mu\nu}+2\alpha\beta g_{\mu\alpha}X^\alpha_{\;\;\;\nu}+\beta^2f_{\mu\nu}.
\end{equation}
In such a case, it is evident that 

\begin{equation}
g_{00}^{eff}dM^2+g_{rr}^{eff}dr^2+(g_{0r}^{eff})(dr dt+dt dr)+r^2d\Omega^2\neq g_{00}dt^2+g_{rr}dr^2+(g_{0r})(dr dt+dt dr)+r^2d\Omega^2,  
\end{equation}
in general. In this sense, then the definition of time $M$ is not essentially the same as the ordinary time definition. This will clearly modify the periodicity structure of the propagator and as a consequence, the radiation coming from the event horizon itself. In addition, the event horizon itself will change. This particular case is more explicit than the standard one analyzed in this manuscript previously. The detailed analysis of this case will be developed in a coming paper.    

\section{Conclusions}   \label{eq:Con}

In this manuscript, I explained the particle creation process of black-holes in dRGT non-linear massive gravity by using the path integral formulation. The first attempt for explaining the particle creation process in dRGT, was done in \cite{1} by the author. In this manuscript, the author focused in the asymptotically flat case with $\beta=(3/4)\alpha^2$. This case represents the situation of zero cosmological constant with a non-zero graviton mass. There is however, a non-trivial configuration of St\"uckelberg fields around the black-hole. They are contained inside the St\"uckelberg function $T_0(r,t)$. By using the St\"uckelberg trick, it was possible to extend the notion of Kruskal coordinates in order to analyze the periodicity properties of the propagator. It was demonstrated that for the observers defining the time in agreement with the St\"uckelberg function $T_0(r,t)$, the periodicity properties of the propagator are the same as in the GR case, then these special observers will not perceive any difference between GR and dRGT at the moment of measuring the temperature. On the other hand, observers defining the time in agreement with $t$, will be able to perceive some extra-component of radiation coming from the vacuum degeneracy reproduced by the extra-degrees of freedom. If the extra-degrees of freedom are relevant after the Vainshtein scale, then only the observers located at scales larger than $r_V$ will be able to perceive the new effects. These observers will experience one of the following situations: 1). Branch point conditions which for the cases of $\alpha\to 0$ or $\alpha\to -2$, suppress any particle creation process. For other cases, with $\alpha$ arbitrary, the perception of particles is affected due to the distortion on the period of the propagator. 2). Shift effect for the case $S=1$ $(\alpha\to\infty)$. In this case, although the period of the propagator is the same with respect to the GR case for any observer, the instants at which the particles appear are different if observers working under the theory of GR compare their results with the ones obtained by observers working under the theory of dRGT. All these effects come from the fact that the extra-degrees of freedom affect the notion of vacuum and as a consequence, the concept of particle. These effects are not related to real emissions of the black-hole event horizon, but rather to the vacuum distortion. The distortion of vacuum inside this theory has been explored in \cite{Higgs}. Different situation appears if the fiducial metric itself is couple to matter. In such a case, it is possible to define a composite or effective metric. It is evident that an effective metric well defined will provide different results with respect to the GR case because we are basically describing a completely different black-hole in such a case. As a consequence, a different horizon and different rates of emission coming from the horizon itself. 
\\

{\bf Acknowledgement}
The author would like to thank Gregory Gabadadze for the kind invitation to New York University (NYU), for the support and discussions around this topic. The author also would like to thank Henry Tye and Jan Hajer from Hong Kong University of Science and Technology, for the kind invitation in order to give a talk in their group, for the support and the subsequent discussions around this topic. The author would like to thank Nobuyoshi Ohta from Kinki University in Osaka for the kind invitation for giving a talk in his group, the support and the subsequent discussions around this topic. Finally the author would like to thank Cosimo Bambi for the invitation to stay during some period at Fudan University in Shanghai, where the author developed part of this project. I.A is supported by the CAS PIFI program.

\end{document}